# Allegation of scientific misconduct increases Twitter attention

Lutz Bornmann* & Robin Haunschild**

*First author and corresponding author:

Division for Science and Innovation Studies

Administrative Headquarters of the Max Planck Society

Hofgartenstr. 8,

80539 Munich, Germany.

E-mail: bornmann@gv.mpg.de

**Contributing author:

Max Planck Institute for Solid State Research

Heisenbergstr. 1,

70569 Stuttgart, Germany.

Email: R.Haunschild@fkf.mpg.de

**Abstract**

The web-based microblogging system Twitter is a very popular altmetrics source for measuring the broader impact of science. In this case study, we demonstrate how problematic the use of Twitter data for research evaluation can be, even though the aspiration of measurement is degraded from impact to attention measurement. We collected the Twitter data for the paper published by Yamamizu et al. (2017). An investigative committee found that the main figures in the paper are fraudulent.

# 1    Introduction

In recent year, alternative metrics (altmetrics) have been introduced as an alternative form of impact measurement in research evaluation; the traditional form of impact measurement is based on citation counts. For altmetric indicators, several sources of meta-data are used for single papers. Most of the sources are known from the social media context: for example, number of tweets or mentions on Facebook (Pooladian & Borrego, 2016). According to Wilsdon et al. (2015) the term altmetrics "is variously used to mean 'alternative metrics' or 'article level metrics', and it encompasses webometrics, or cybermetrics, which measure the features and relationships of online items, such as websites and log files. The rise of new social media has created an additional stream of work under the label altmetrics" (pp. 5-6). Although research on altmetrics is a relatively new field in scientometrics, several literature overviews have already been published. The most recent and comprehensive overview is from Sugimoto, Work, Larivière, and Haustein (2016).

One of the most important factors for the popularity of altmetrics is its possible use for the broad impact measurements of research (Bornmann, 2014; Haustein, 2014). Traditional metrics focus on the science sector; thus, they cannot fulfill the demands from governments and funding organizations to measure the impact of research beyond science: "Governments and funding organizations are increasingly asking scholars to demonstrate societal impact and relevance, in addition to scientific excellence … Altmetrics … has been advocated as a potential indicator of such impact" (Sugimoto et al., 2016). The broader impact measurement is especially relevant in the Research Excellence Framework (REF); the Higher Education Funding Council for England (HEFCE) considers different types of impact inside and outside of academia (Mohammadi, Thelwall, & Kousha, 2016).

The web-based microblogging system Twitter is a very popular altmetrics source for measuring the broader impact of science (Mas-Bleda & Thelwall, 2016). Many scientometric studies on altmetrics have used Twitter data and many publishers add Twitter information to their published papers – reflecting the attention a paper receives. Twitter data are also listed in the Snowball Metrics Recipe Book (Colledge, 2014) – an initiative of several universities to standardize the set of indicators for research evaluation purposes. In this case study, we would like to demonstrate how problematic the use of Twitter data for research evaluation can be, even though the aspiration of measurement is degraded from impact to attention measurement, such as by Moed (2017): "the current author agrees with the proposition that usage-based indicators and altmetrics primarily reflect attention rather than influence" (p. 133).

## 2 Methods and results

In March 2017, the paper "In Vitro Modeling of Blood-Brain Barrier with Human iPSC-Derived Endothelial Cells, Pericytes, Neurons, and Astrocytes via Notch Signaling" was published by Yamamizu et al. (2017) in the journal *Stem Cell Reports*. A full investigative committee found "that all six main figures in the paper were fraudulent. It also concluded that the lead author, Kohei Yamamizu, an assistant professor at CiRA, had carried out the fabrications on his own. The images are central to the paper's conclusions, and the authors have asked the journal to retract the paper" (see http://www.sciencemag.org/news/2018/01/nobel-laureate-suggests-he-could-resign-leadership-post-over-colleague-s-bogus-paper).

We collected the Twitter data for the paper published by Yamamizu et al. (2017) from the Altmetric.com API on January 26, 2018. Figure 1 shows the number of tweets per day. The paper stimulated some online activity (42 tweets) shortly after appearance, but 502 tweets followed when the misconduct discussion around the paper started. The bulk of the 544 tweets mentioning this paper appeared on only two days: 220 tweets appeared on 22 January 2018 and 235 tweets the following day. Measured by the Attention Score (date of search 26

January 2018), the paper is "in the top 5% of all research outputs scored by Altmetric" (see https://www.altmetric.com/details/16752529/twitter). Thus, the altmetric indicators point out that the paper by Yamamizu et al. (2017) produced massive attention in science and beyond. The non-expert user of altmetrics data for the paper could conclude – if the data are used as a proxy for societal impact – that the paper is a useful contribution from science for the public.

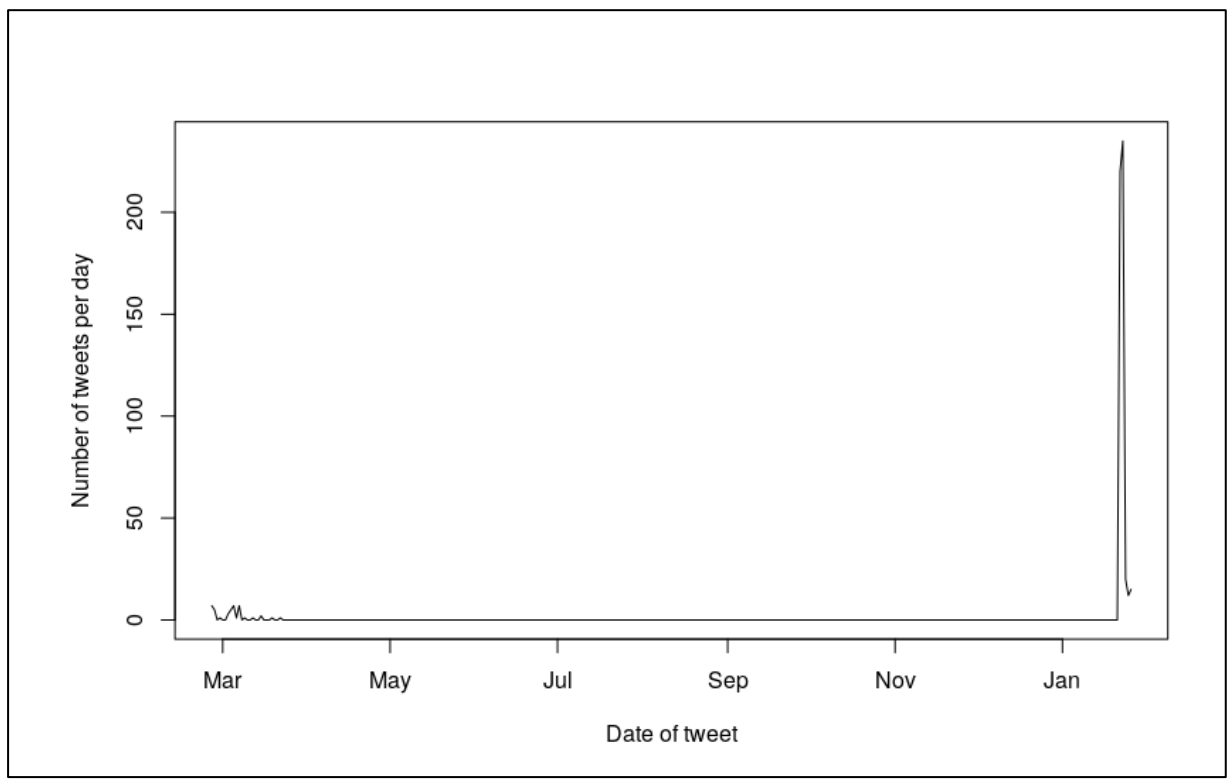

Figure 1. Number of tweets per day for the paper by Yamamizu et al. (2017)

## 3  Discussion

The result of this case study clearly reveals the problems with using Twitter data for measuring impact or attention of science. Without considering the content of the tweets, simple counting can lead to wrong conclusions. However, cases of scientific misconduct are not the only "outliers" which distort the validity of altmetrics data; Sugimoto (2016) mentions some others: "One can easily find examples of extremely high Altmetric.com scores which are the result of a viral joke, proofreading error, or scientific hoax. Behind these outliers are undoubtedly scores of articles whose recognition in policy documents, popular press, and on social media is a legitimate sign that the work is relevant and interesting to a broader public. How to identify the underlying mechanism of altmetric attention remains a critical challenge".

The result of this study is in agreement with the critical view of Robinson-Garcia, Costas, Isett, Melkers, and Hicks (2017) on Twitter data. The authors analyzed comprehensive altmetrics data for papers published in journals from the field of dentistry. They question the use of altmetrics data, because tweets do not (cannot) reflect an intellectual analysis with the paper in question: "A multi-year campaign has sought to convince us that counting the number of tweets about papers has value. Yet, reading tweets about dental journal articles suggested the opposite. This analysis found: obsessive single issue tweeting, duplicate tweeting from many accounts presumably under centralized professional management, bots, and much presumably human tweeting duplicative, almost entirely mechanical and devoid of original thought". The authors conclude that tweets should not be used for research evaluation.

Proponents of altmetrics might oppose that cases of misconduct will not only have high Twitter counts, but also high citation counts. Thus, citations are concerned with the same problem as tweets. However, there is a decisive difference between both data sources: tweets increase with the detection of misconduct – the misconduct boosts the attention – which is not the case with citations. The results of the study by Shuai et al. (2017) show that citation impact decreases when the paper in question has been retracted: "the scholarly impact of retracted papers and authors significantly decreases after retraction, and the most severe impact decrease correlates with retractions based on proven, purposeful scientific misconduct" (p. 2225).

However, we do not want to throw the baby out with the bath water. Social media (including Twitter) are useful instruments to draw attention to cases of scientific misconduct (Gross, 2016). Thus, scientists are better informed on these cases by using social media; only simple counting of tweets for research evaluation purposes should be avoided.